Human Hunting Evolved as an Adapted Result of Arboreal Locomotion Model of Two-arm Brachiation (II)

C.Fang[1], T.Jiang[2]

[1]Department of Engineering Mechanics, Chongqing University, Chongqing, 400044, China.

[2]College of Resources and Environmental Science, Chongqing University, Chongqing, 400044, China.

*Corresponding author: C.Fang

**Various fossil evidences show that hunting is one of major means of ancient human to get foods [1-3]. But our ancestors' running speed was much slower than quadruped animals, and they did not have sharp claws and canines. So, they have to rely heavily on stone and wooden tools when they hunting or fighting against other predators, which are very different from the hunting behaviors of other carnivores. There are mainly two types of attack and defense action during human hunting, front or side hit with a wooden stick in hands and stone or wooden spears throwing (Fig.1), and throwing had play an important role in human evolution process. But there is almost no work to study the why only human chose to hunting by this way. Here we suppose that ancient human chose two-arm brachiation as main arboreal locomotion mode because of their suitable body weight. Human body traits include slim body, parallel arranged scapulas, long thumb and powerful grip ability are all evolved as results of two-arm brachiation. The relevant adaptive evolution of the shoulder bone structure make human arms with a large range of movement and the long thumb makes human activities to be more accurate and controllable. These are two important body structure advantages of ancient human which makes them could get from arboreal life into a whole new hunting and fighting stage.**

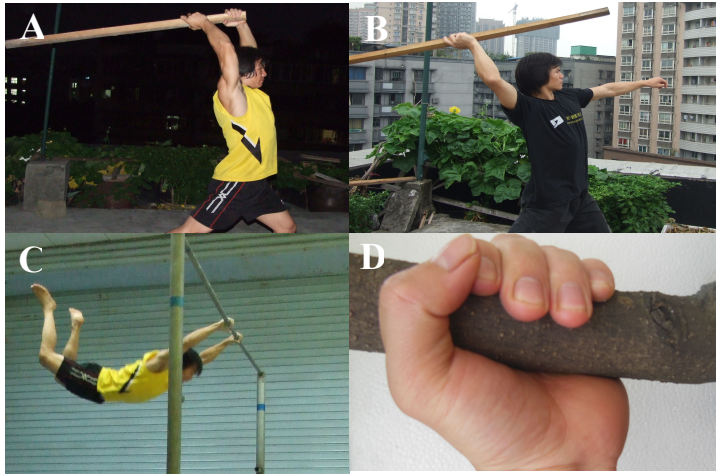

Fig.1. Flexible shoulder joint and long thumb are the key biological structures for human hunting (A, B), this two traits are closely related to arboreal locomotion of two-arm brachiation(C, D).

To performing the hitting and throwing actions well, at first, human shoulder joints should have excellent mobility to generate high speed and power [4, 5], therefore, the two scapulas must to be arranged in parallel in stead of oblique. Parallel arranged scapulas is indeed a very unique biological advantages of human, even chimpanzees can not throw stones so powerfully and skillfully like us. But when did human scapula evolution happen? If this took place after human leave from the forest and evolved slowly by the need of hunting, then there would be a long period of hard transition time because of their relatively small body and poor throwing ability. Therefore, we speculate that this process may happen before they leave forest, and accomplished relatively quickly, but how?

Secondly, since early humans were slow and small, their survival depends highly on the use of natural tools, and human's strong thumb and its special structure makes the use of tools became very convenient[6,7]. Only human's thumb can help to powerfully and precisely control many kinds of tools, this is our another outstanding features, if our ancestors could control the throwing speed, direction, distant of the stones and spears more precisely, then they have more chance to survival. And the intelligence of human developed along with the skillfully use and precisely control of a variety of tools.

For the same reason，we think this function of human thumb could not be evolved after arboreal period. Then, how was human's strong thumb evolved? We attempt to explore these two questions and analyze the evolution mechanism by mechanics.

**1 It's reasonable for arboreal human move by two-arm brachiation because of their body weight.**

Most of animal's locomotion models, for example, running, jumping, flying and arm swing have to overcome self-weight. Animas' body weight always determines their suitable locomotion model. Cross arm swing is one of major locomotion model of apes because their weights are greater than the general monkeys, and this locomotion made the body of apes very different from that of monkeys.

For our arboreal ancestors, arm swing should be one major locomotion model. But since all apes have their own arm swing patterns, which one was most suitable for our ancestors? If we assume their body weight is about 40KG, the distance between branches to the body gravity center was one meter, and the swing speed was 6 meters per second, then the tension force of the upper limb is 1760N. It's very hard to bear this tension force by only one arm, so here we assume our ancestors chose to take two-arm brachiation instead of the cross arm swing model which favored by gibbons.

**2 Human scapula evolution and two-arm brachiation**

The shape and arrangement of scapula of chimpanzees and that of human are significantly different (Fig.2). The two scapulas of chimpanzee are oblique for the need of quadruped walking, but that of humans were evolved to be arranged paralleled. We called this change of human shoulder "flattening process" and it enable human a great range of the independence movement of upper limbers, and along this change, human could get much more hit speed and power which enhance their survivability greatly. Therefore, we speculate that human scapulas evolved from oblique to flatten state rapidly and eventually arranged in parallel, and its direct driving force is the gravity and centrifugal force of two-arm branchiation during the arboreal period of human.

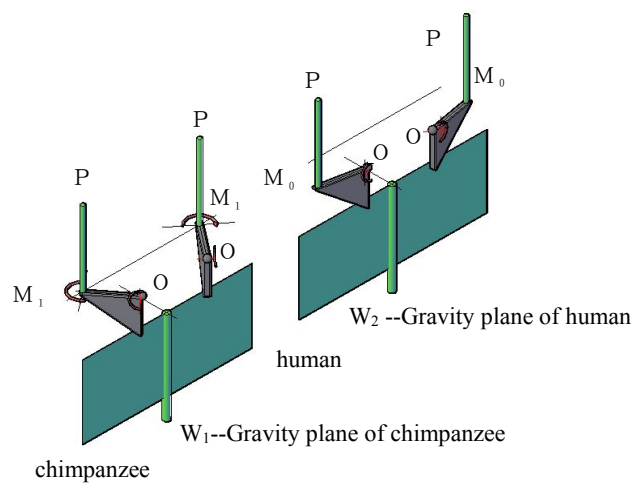

Fig.2  Mechanic analysis of two-arm hang of chimpanzee and human

If we assume that the angle of the scapulas of earlier arboreal human are similar with that of modern apes (Fig.2 Left). P is the suspension tension stress of the arms and $W_1$ and $W_2$ are the gravity planes of ape and human. Figure 3 and Figure 4 is top view of two-arm suspension, if the two scapulas were oblique with each other, then stress P would not be in the gravity plane $W_1$, and inevitably produce a torque $M_1$, $M_1$ make the scapulas rotate toward to the plane $W_1$, until the scapulas are almost in the gravity plane, then the torque $M_1$ equal to zero, this is the last balance mechanic state. If we considered the rotation of the body around the branches, than we have to count the effect of centrifugal force also.

$M_1$ = the horizontal component of gravity and centrifugal force * arm (specific calculation)

= $(G+F_c)*L$

Therefore, the gravity and centrifugal force made the human body became very slender, and the rotational torque $M_1$ caused by the join forces were the directed causes of parallel arranged scapulas. After the flattening of human scapulas, the upper limbs could move back-forward very freely and it's efficient for the strength of the trunk pass through the scapula to upper limbs. This can explain that why human shoulders are so flexible and the hitting and throwing ability of human are much more powerful than that of chimpanzees.

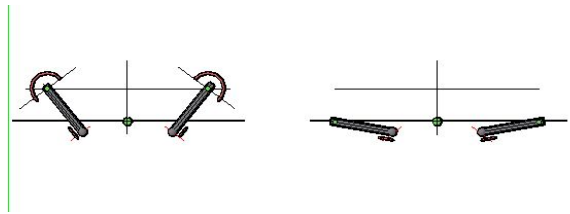

Fig.3 Top view of the scapula of chimpanzee (left) and human (right)

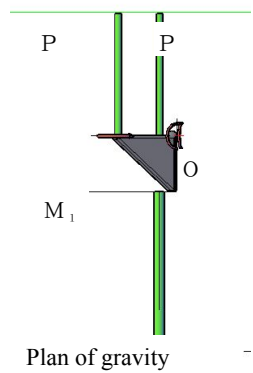

Fig.4 Mechanical Analysis of scapula bone when hanging by two arms

**3 Thumbs must participate in two-arm brachiation**

The main function of the human thumb is clamp objects with other four fingers, so we can take many objects stably and accurately. It is because of the maximizing of this function, human can most effectively use many kinds of tools, which is also one of the main differences between human and apes. The thumbs of Australopithecus sediba were even significantly longer than that of modern human [8]. So, when and how our long and strong thumbs evolved? Previous studies did not give a clear answer. Some researchers speculate it was gradually evolved by the use of tools after leaving the forest. But this hypothesis means chimpanzees could evolve long thumbs by the same process also, it is unacceptable. So we assume that the evolution of human thumb also have completed during the arboreal period.

Evolution is always closely related with the change of the survival manner, we could reasonably explain the evolution of human thumb by arboreal two-arm branchiation. During two-arm swing, the tension stress are mainly rely on the four fingers, but when the body are heavy and swing speed are high, the suitable branches must be thick to withstand the composition of gravity and centrifugal forces, otherwise the branch will break. According to the experiment of modern adult to grasp branches (Figure 5), if the branch diameter is less than 30mm (Figure 5a), four fingers could also hold the branch, but far less stable than the manner of grip with thumb (Figure 5b). If the branch diameter is more than 50mm (Figure 5c and 5d), we can strongly and stably grasp the thick branches only by the participation of thumbs, it's dangerous and hard to hold the branches without thumbs.

Most importantly, if they have to swing to some higher branches by two-arm brachiation, their body will be in a hypsokinesis position, which is dangerous and imbalance, and they have to turning their bodies back to normal state by push the branches to get a torque before taking off. They could get much greater torque if their thumbs and palm oppositely grip the branches. This mechanical mechanism could explain why human have long thumb and short palm.

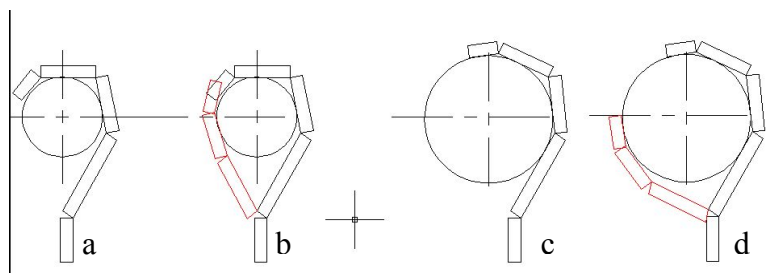

Fig.5. Comparison of two manners to grip the fine and thick branches

Due to the body weight condition, ancient human had to grasp the thick branches to swing through forest quickly, so our arboreal ancestors had already evolved strong thumbs to significantly enhance the grip strength. Moreover, with the participate of thumbs, the palm, fingers could get as much contact areas as possible with branches which makes the hands and branch fixed as a whole, this ensure the stable and accurate of the whole process of two-arm branchiation, for example, positioning, starting, flying, stop and other controlled and skilled movements. And these controlled and skilled movements maybe closely related to the evolution of human intelligence. Therefore, our hypothesis is that human thumb was evolved in the forest, together with the flattening of scapulas, were two pre-prepared physical advantage for the survival of terrestrial life afterward.

**4 Conclusions**

Because of the suitable body weight and other unknown factors, ancient human learned to swing by both two arms, and from the view of mechanics, this locomotion of two-arm branchiation can reasonably interpreted the evolution of the human scapulas and thumbs, and human's unique scapulas and thumbs had laid a biological basis for the leaving of the forest and stating a hunter-gatherer living.

**Supplementary Materials:**

Movies S1-S3

Respectively, from website:

http://v.ku6.com/show/3eOJMLiL_wyrm-l2aoX1sg...html

http://v.ku6.com/show/gTMAqNDtfsQGK0yF0_Bjqg...html

http://video.sina.com.cn/v/b/113861422-3735632923.html